\documentclass[12pt]{article}
\usepackage{epsfig}
\usepackage{amssymb}
\usepackage{amsmath}
\usepackage{amsfonts}
\usepackage{graphicx}
\usepackage{mathrsfs}
\DeclareMathAlphabet{\mathscrbf}{OMS}{mdugm}{b}{n}
\usepackage[dvips]{color}
\usepackage{multirow}
\usepackage{calc}
\usepackage{accents}

\makeatletter
\newcommand{\shorteq}{\mathrel{\mkern0.2mu\mathpalette\shorteq@\relax\mkern0.2mu}}
\newcommand{\shorteq@}[2]{\scalebox{0.5}[1]{$\m@th#1=$}}

\newcommand{\longeq}[1]{\mathrel{\mathpalette\longeq@{#1}}}
\newcommand{\longeq@}[2]{%
  \begingroup
  \sbox\z@{$\m@th#1=$}%
  \ifdim#2<\wd\z@
    \resizebox{#2}{\height}{\box\z@}%
  \else
    \ifdim#2<3\wd\z@
      \hbox to #2{$\m@th#1=\hss=\hss=\hss=$}%
    \else
      \hbox to #2{$\m@th#1=\cleaders\hbox to 0.2\wd\z@{\hss$#1=$\hss}\hfil=$}%
    \fi
  \fi
  \endgroup
}
\makeatother

\newcommand{\bsigma}{\boldsymbol{\sigma}}

\newcommand{\bOmega}{\boldsymbol{\Omega}}

\newcommand{\A}{\mathbb{A}}
\newcommand{\R}{\mathbb{R}}
\newcommand{\C}{\mathbb{C}}
\newcommand{\Z}{\mathbb{Z}}

\newcommand{\ff}{\mathfrak{f}}

\newcommand{\fK}{\mathfrak{K}}

\newcommand{\fT}{\mathfrak{T}}

\newcommand{\bk}{\mathbf{k}}

\newcommand{\bn}{{\mathbf{n}}}
\newcommand{\bfr}{\mathbf{r}}

\newcommand{\bx}{\mathbf{x}}

\newcommand{\bu}{\mathbf{u}}

\newcommand{\bF}{\mathbf{F}}

\newcommand{\bH}{\mathbf{H}}
\newcommand{\bI}{\mathbf{I}}

\newcommand{\bL}{\mathbf{L}}

\newcommand{\bM}{\mathbf{M}}

\newcommand{\bS}{\mathbf{S}}

\newcommand{\bU}{\mathbf{U}}

\newcommand{\cH}{\mathcal{H}}

\newcommand{\cK}{\mathcal{K}}

\newcommand{\cP}{\mathcal{P}}

\newcommand{\cT}{\mathcal{T}}
\newcommand{\cU}{\mathcal{U}}

\newcommand{\cX}{\mathcal{X}}

\newcommand{\be}{\begin{equation}}
\newcommand{\ee}{\end{equation}}
\newcommand{\bea}{\begin{eqnarray}}
\newcommand{\eea}{\end{eqnarray}}
\newcommand{\nn}{\nonumber}
\newcommand{\kt}{\rangle}
\newcommand{\br}{\langle}

\newcommand{\ed}{\end{document}}

\newcommand{\rn}{{\rm n}}

\newcommand{\ru}{{\rm u}}
\newcommand{\rk}{{\rm k}}

\newcommand{\rr}{{\rm r}}

\newcommand{\bi}{\begin{itemize}}
\newcommand{\ei}{\end{itemize}}

\newcommand{\bce}{\begin{center}}
\newcommand{\ece}{\end{center}}

\newcommand{\sD}{\mathscr{D}}

\newcommand{\sF}{\mathscr{F}}

\newcommand{\sH}{\mathscr{H}}

\newcommand{\sT}{\mathscr{T}}
\newcommand{\sV}{\mathscr{V}}

\newcommand{\RE}{{\rm Re}}

\newcommand{\bcK}{{\boldsymbol{\cK}}}

\newcommand{\bcH}{{\boldsymbol{\cH}}}
\newcommand{\bcU}{{\boldsymbol{\cU}}}

\newcommand{\for}{{\mbox{\rm for}}}

\begin{document}

%\title{Reciprocity Theorem, Fundamental Transfer Matrix,\\ and Its Anti-Pseudo-Unitarity}

\title{Reciprocity Theorem and Fundamental Transfer Matrix}

\author{Farhang Loran\thanks{E-mail address: loran@iut.ac.ir}
~and Ali~Mostafazadeh\thanks{Corresponding author, E-mail address:
amostafazadeh@ku.edu.tr}\\[6pt]
$^*$Department of Physics, Isfahan University of Technology, \\ Isfahan 84156-83111, Iran\\[6pt]
$^\dagger$Departments of Mathematics and Physics, Ko\c{c} University,\\  34450 Sar{\i}yer,
Istanbul, T\"urkiye}

\date{ }
\maketitle

\begin{abstract}

Stationary potential scattering admits a formulation in terms of the quantum dynamics generated by a non-Hermitian effective Hamiltonian. We use this formulation to give a proof of the reciprocity theorem in two and three dimensions that does not rely on the properties of the scattering operator, Green's functions, or Green's identities. In particular, we identify reciprocity with an operator identity satisfied by an integral operator $\widehat{\mathbf{M}}$, called the fundamental transfer matrix. This is a multi-dimensional generalization of the transfer matrix $\mathbf{M}$ of potential scattering in one dimension that stores the information about the scattering amplitude of the potential. We use the property of $\widehat{\mathbf{M}}$ that is responsible for reciprocity to identify the analog of the relation, $\det{\mathbf{M}}=1$, in two and three dimensions, and establish a generic anti-pseudo-Hermiticity of the scattering operator. Our results apply for both real and complex potentials.

%\noindent PACS numbers: 03.65.Nk, 42.25.Bs\vspace{2mm}

%\noindent Keywords: Scattering, transfer matrix in two dimensions, delta-function potential supported on a line, arrays of delta-functions, spectral singularity

\end{abstract}

\section{Introduction}

In one dimension (1D) the time-independent Schr\"odinger equation,
	\be
	-\psi''(x)+v(x)\psi(x)=k^2\psi(x),
	\label{sch-eq-1d}
	\ee
defines a scattering problem for $k\in\R^+$ provided that $v$ is a real or complex short-range potential.\footnote{In $d+1$ dimensions, a function $v:\R^{d+1}\to\C$ is called a short-range potential if $\int_{\R^{d+1}}d^{d+1}x\,|v(\bfr)|<\infty$ and $\lim_{|\bfr|\to\infty}|\bfr|^{d/2+1}v(\bfr)=0$, \cite{yafaev}.} For such a potential, every solution of (\ref{sch-eq-1d}) fulfills 
	\be
	\psi(x)\to A_\pm e^{ikx}+B_\pm e^{-ikx}~~{\rm for}~~x\to\pm\infty,
	\label{pw}
	\ee
where $A_\pm $ and $B_\pm $ are possibly $k$-dependent coefficients \cite{yafaev}. There is a unique $k$-dependent $2\times 2$ matrix $\bM$ that is independent of these coefficients and connects them according to
	\be
	\bM\left[\begin{array}{c} A_-\\ B_-\end{array}\right]=
	\left[\begin{array}{c} A_+\\ B_+\end{array}\right].
	\label{M-1d}
	\ee
This is called the transfer matrix of the potential $v$, \cite{sanchez,tjp-2020}.

The scattering setups corresponding to the source of the incident wave being located at $x=-\infty$ and $x=+\infty$ are respectively described by the left-incident and right-incident solutions, $\psi^l$ and $\psi^r$, of (\ref{sch-eq-1d}) for which (\ref{pw}) takes the form:
	\begin{align}
	&\psi^{l}(x)\to N^{l}\times
	\left\{\begin{array}{ccc}
    	 e^{ikx}+{R}^{l}\,e^{-ikx} & {\rm for}& x\to-\infty,\\
    	 {T}^{l}\,e^{ikx} & {\rm for}& x\to+\infty,
    	\end{array}\right.
	\label{jost-l}\\
    	&\psi^{r}(x)\to N^{r}\times 
	\left\{\begin{array}{ccc}
    	 {T}^{r}\,e^{-ikx} & {\rm for}& x\to-\infty,\\
    	 e^{-ikx} + {R}^{r}\,e^{ikx}  & {\rm for}& x\to+\infty.
    	\end{array}\right.
	\label{jost-r}
	\end{align}
Here $N^{l/r}$ are nonzero complex coefficients representing the amplitude of the incident wave, and ${R}^{{l/r}}$ and ${T}^{{l}/{r}}$ are $k$-dependent coefficients known as the left/right reflection and transmission amplitudes of the potential, respectively.\footnote{The left/right reflection and transmission coefficients are respectively given by $|{R}^{{l/r}}|^2$ and $|{T}^{{l}/{r}}|^2$.}

Comparing \eqref{jost-l} and \eqref{jost-r} with \eqref{pw}, we can express the coefficients $A_\pm$ and $B_\pm$ associated with $\psi^{{l/r}}$ in terms of $N^{l/r}$, $R^{l/r}$, and $T^{l/r}$. Substituting these in (\ref{M-1d}), we find
	\begin{align}
	&R^{l}=-\frac{M_{21}}{M_{22}}, &&T^{l}=\frac{\det\bM}{M_{22}},
    	&&R^{r}=\frac{M_{12}}{M_{22}}, &&T^{r}=\frac{1}{M_{22}},
         \label{RT=}
         %\\
	%&M_{11}={T}^{l}-\frac{{R}^{l}{R}^{r}}{{T}^{r}},
	%&&M_{12}=\frac{{R}^{r}}{{T}^{r}},
	%&&M_{21}=-\frac{{R}^{l}}{{T}^{r}},
	%&& M_{22}=\frac{1}{{T}^{r}},
    	%\label{Mij=}
    	\end{align}
where $M_{ij}$ stand for the entries of $\bM$, \cite{tjp-2020}. According to \eqref{RT=}, the solution of the scattering problem for the potential $v(x)$, which means the determination of the reflection and transmission amplitudes, is equivalent to finding the transfer matrix.

Let $\psi_1$ and $\psi_2$ be any pair of solutions of (\ref{sch-eq-1d}), and $W$ denote their Wronskian, i.e.,
   \be
    W(x):=\psi_1(x)\psi_2'(x)-\psi_2(x)\psi_1'(x).
    \label{Wronskian}
    \ee
Then (\ref{sch-eq-1d}) implies $W'(x)=0$, i.e., $W$ does not depend on $x$. If we compute the Wronskian of $\psi^{{l}}$ and $\psi^{{r}}$, and use (\ref{jost-l}) and \eqref{jost-r} to compute its value for $x\to-\infty$ and $x\to+\infty$, we obtain $2ik/T^{l}$ and $2ik/T^{r}$, respectively \cite{tjp-2020}. Therefore, 
	\be
	T^{l}=T^{r}.
	\label{T=T1}
	\ee
This proves the following reciprocity theorem in one dimension.
	\begin{itemize}
	\item[]{\bf Theorem~1 (Reciprocity in 1D)} {\em Let $v:\R\to\C$ be a {real or complex} short-range potential. Then its left and right transmission amplitudes coincide.} 
	\end{itemize}
It is important to note that this theorem applies to real as well as complex short-range potentials.\footnote{The claim, made for example in \cite{inv-scat}, that reciprocity is a consequence of unitarity is false.} In light of \eqref{RT=}, it implies
	\be
	T^{l}=\frac{1}{M_{22}}.
	\label{Reciprocity-1D}
	\ee
We can also view it as a consequence of a property of the transfer matrix, namely 
	\be
	\det\bM=1.
	\label{det-M}
	\ee
We can establish this relation using a curious link between the transfer matrix and the evolution operator for the two-level quantum system defined by the time-dependent non-Hermitian matrix Hamiltonian:
	\be
	\bcH(t):=\frac{v(t)}{2k}\left[\begin{array}{cc}
	1 & e^{-2ikt}\\
	-e^{2ikt} & -1\end{array}\right].
	\label{H-1D}
	\ee
Note that on the right-hand side of this relation, the time variable $t$ is substituted for $x$ in $v(x)$. Therefore, it is the space coordinate $x$ that plays the role of time. This does not mean that $v$ is a time-dependent potential. Indeed, $t$ is an effective time parameter that has the physical meaning of the space coordinate $x$. To emphasize this point, in the following we use $x$ as an evolution parameter, i.e., set $t:=x$.
	
Let $x_0$ be an arbitrary initial effective time, and $\bcU(x,x_0)$ denote the evolution operator corresponding to the Hamiltonian \eqref{H-1D}, i.e., $\bcU(x,x_0)$ is the unique solution of	
	\begin{align}
	i\partial_x\bcU(x,x_0)=\bcH(x)\,\bcU(x,x_0),~~~~~~~\bcU(x_0,x_0)=\bI,
	\label{sch-eq-u}
	\end{align}
where $\bI$ is the $2\times 2$ identity matrix. Then, as shown in Refs.~\cite{tjp-2020,ap-2014},
	\be
	\bM=\bcU(+\infty,-\infty)=\sT\exp\left[-i\int_{-\infty}^\infty dx\:\bcH(x)\right],
	\label{M=1D}
	\ee
where $\sT$ is the time-ordering operator, so that
	\begin{align}
	\bcU(x,x_0)&=\sT~\exp\left[-i\int_{x_0}^x\! dx'\:\bcH(x')\right]\nn\\
	&:=\bI+\sum_{n=1}^\infty (-i)^n\int_{x_0}^x\!\! dx_n\int_{x_0}^{x_n}\!\! dx_{n-1}
	\cdots\int_{x_0}^{x_2}\!\! dx_1\:\bcH(x_n)\bcH(x_{n-1})\cdots \bcH(x_1).
	\nn
	\end{align}
Because $\bcH(x)$ is traceless, \eqref{sch-eq-u} implies that  $\det\bcU(x,x_0)=1$. In particular, \eqref{det-M} holds. In view of \eqref{RT=}, this provides an alternative proof of Theorem~1.

A three-dimensional (3D) generalization of Theorem~1 was discovered by Helmholtz in his studies of sound waves, perfected by Lord Rayleigh, and extended to the scattering of electromagnetic waves by Lorentz.\footnote{For a brief history of the reciprocity theorem and its applications in fluid dynamics, see Refs.~\cite{Masoud-2019} and \cite{Ramin}.} Various definitions of reciprocity has been considered in Ref.~\cite{potton} and the study of their origins and implications continues to attract much attention. For instance a reciprocity theorem for the two-dimensional Helmholtz equation with boundary conditions is given in Ref.~\cite{RMP2022}. See also Ref.~\cite{Twersky1954}. Additionally, reciprocity theorems for scattering in two dimensional (2D) time-dependent materials are formulated in Ref.~\cite{Wapenaar}.  In Ref.~\cite{Deak-Fulop}, the reciprocity theorems for scalar and vector waves are linked with the existence of a reciprocity operator. This is an antiunitary operator $\widehat\fK$ that commutes with the free Hamiltonian operator $\widehat H_0$ and satisfies 
	\be
	\widehat v^\dag=\widehat \fK\,\widehat v\,\widehat \fK^{-1},
	\label{anti-ps-v}
	\ee 
where $\widehat v$ is the operator representing the interaction potential, so that the Hamiltonian operator that defines the scattering problem through the time-independent Schr\"odinger equation has the form $\widehat H=\widehat H_0+\widehat v$. See also \cite{Sigwarth}.  Because $\widehat H_0$ commutes with $\widehat\fK$, \eqref{anti-ps-v} is equivalent to 
	\be
	\widehat H^\dag=\widehat \fK\,\widehat H\,\widehat \fK^{-1}.
	\label{anti-ps-H}
	\ee 
If $\widehat \fK$ happens to be Hermitian, which is the case if and only if it is an involution, \eqref{anti-ps-H} states that $\widehat H$ is $\widehat\fK$-anti-pseudo-Hermitian \cite{p3,paper183}.\footnote{If $\widehat\fK$ is not Hermitian, \eqref{anti-ps-H} implies that $\widehat\fK^2$ is a linear unitary operator that commutes with $\widehat H$. This means that it is either a function of $\widehat H$ or corresponds to a symmetry transformation.}

Given a possibly complex-valued short-range potential $v:\R^{d+1}\to\C$ with $d\in\{1,2\}$, the scattering solutions of the Schr\"odinger equation,
    \be
    -\boldsymbol{\nabla}^2\,\psi(\bfr)+v(\bfr)\,\psi(\bfr)= k^2\,\psi(\bfr),
    \label{sch-eq}
    \ee
satisfy
	\be
    	\psi(\bfr)\to\frac{N}{(2\pi)^{\frac{d+1}{2}}}\left[e^{i\bk_0\cdot\bfr}+ 
	\frac{e^{ikr}}{r^{\frac{d}{2}}}\,\ff(\bn_0,\bn)\right]~~~\for~~~r\to\infty,
	\label{asymptot}
 	\ee
where $k$ and $N$ are respectively the wavenumber and amplitude of the incident wave, $r:=\left|\bfr\right|$, $\ff(\bn_0,\bn)$ is the scattering amplitude, and $\bn_0$ and $\bn$ are respectively the unit vectors along the incident and scattered wave vectors, $\bk_0$ and $\bk:=k r^{-1}\bfr$, i.e., 
    \begin{align}
    &\bn_0:=k^{-1}\bk_0,
    &&\bn:=k^{-1}\bk=r^{-1}\bfr.\nn
    \end{align}
    
Landau and Lifshitz \cite{Landau} state and prove the following reciprocity theorem for the special case where $d=2$ and $v$ is a real potential.
	\begin{itemize}
	\item[]{\bf Theorem~2 (Reciprocity in 2D and 3D)} {\em Let $d\in\{1,2\}$ and $v:\R^{d+1}\to\C$ be a {real or complex} short-range potential. Then the scattering amplitude $\ff(\bn_0,\bn)$ of $v$ satisfies}
	\be
	\ff(\bn_0,\bn)=\ff(-\bn,-\bn_0).
	\label{reciprocity-D23}
	\ee 
	\end{itemize}
	
The link between the transfer matrix $\bM$ of a short-range potential and the dynamics generated by the effective non-Hermitian Hamiltonian operator \eqref{H-1D} in 1D admits 2D and 3D generalizations. This provides the basis of a dynamical formulation of stationary scattering in which the scattering amplitude of the potential is extracted from a generalization of the transfer matrix which we call the fundamental transfer matrix \cite{pra-2021}. This is a $2\times 2$ matrix $\widehat\bM$ whose entries $\widehat M_{ij}$ are linear (integral) operators acting in an infinite-dimensional function space. Similarly to $\bM$, the fundamental transfer matrix admits an expression in terms of the evolution operator for an affective non-Hermitian Hamiltonian operator.

The main purpose of the present article is to unravel the basic property of $\widehat\bM$ that is responsible for the reciprocity relation~\eqref{reciprocity-D23}. In particular, we provide 2D and 3D generalizations of Eqs.~\eqref{RT=}, \eqref{Reciprocity-1D}, and \eqref{det-M}, and give a proof of Theorem~2 that relies on an operator identity satisfied by $\widehat\bM$.

\section{Dynamical formulation of stationary scattering in 2D and 3D}
\label{Sec2}

\begin{figure}
    \begin{center}
        \includegraphics[scale=.25]{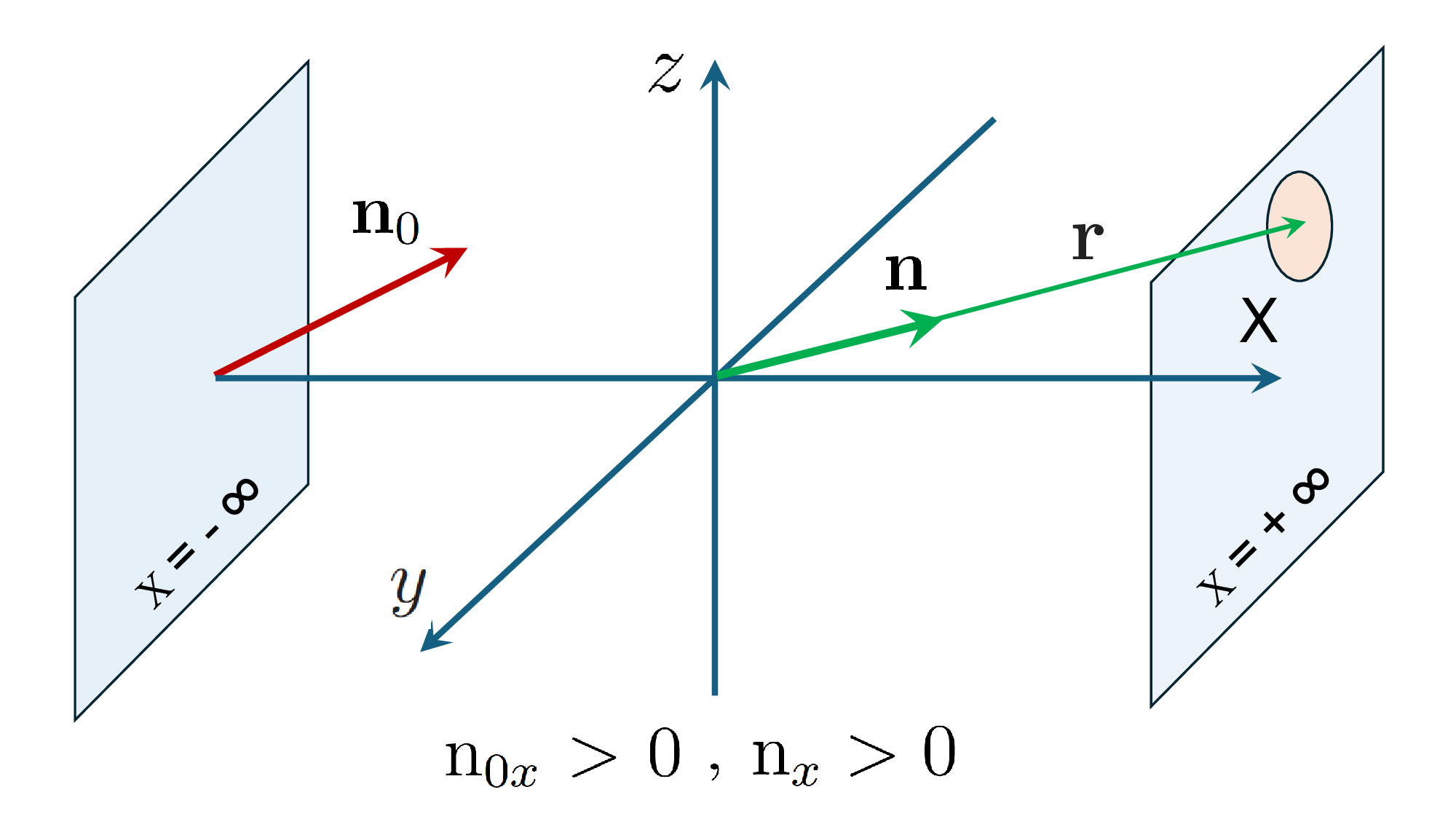}~~
        \includegraphics[scale=.25]{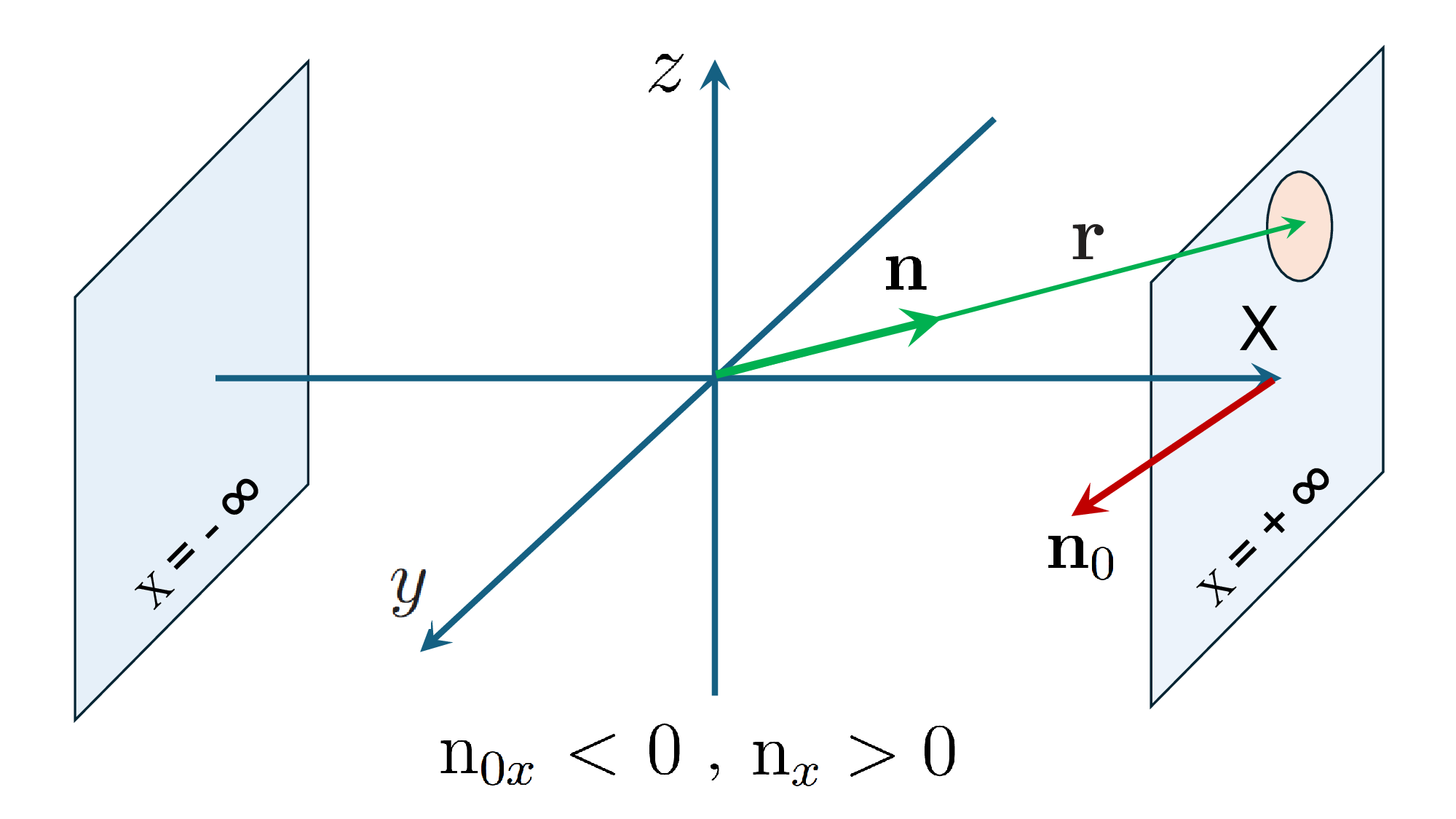}
        \caption{Schematic view of the scattering setup for a
        left-incident wave (on the left) and a right-incident
        wave (on the right). $\bn_0$ and $\bn$ are respectively the unit vectors along the incident and scattered wave vectors. For the left- and
        right-incident waves, the $x$ component of $\bn_{0}$ is respectively positive and negative. The orange elliptic region represents a detector screen position at $x=+\infty$. This corresponds to $\bn$ having a positive $x$ component.}
        \label{fig1}
    \end{center}
    \end{figure}

Consider the scattering problem defined by the Schr\"odinger equation~\eqref{sch-eq} in 3D. Then the symbol $\bfr$ appearing in \eqref{asymptot} represents the position of a generic detector placed at spatial infinity. Without loss of generality, we can imagine that the detectors measuring the scattered wave are located on the planes $x=\pm\infty$, where $x,y$, and $z$ are Cartesian coordinates of $\bfr$, and the source of the incident wave reside on either of the planes, $x=-\infty$ and $x=+\infty$. We call the corresponding incident waves ``left-'' and ``right-incident waves,'' respectively. If we use $\rn_{0x}$ and $\rn_x$ to denote the $x$ components of $\bn_0$ and $\bn$, then for left- and right-incident waves, $\rn_{0x}>0$ and  $\rn_{0x}<0$, respectively. Similarly, for detectors positioned at $x=-\infty$ and $x=+\infty$, we have $\rn_x<0$ and  $\rn_x>0$, respectively. Figure~\ref{fig1} provides a schematic view of this scattering setup with a detector placed at $x=+\infty$.

{Next, we introduce a useful notation: For each $\bu:=(\ru_x,\ru_y,\ru_z)\in\R^3$,  we use $\vec\ru$ to denote $(\ru_y,\ru_z)$. This allows us to identify $\bu$ with $(\ru_x,\vec\ru)$. We refer to $\vec\ru$ as the projection of $\bu$ onto the $y$-$z$ plane.}

The above discussion applies to the scattering problem defined by the Schr\"odinger equation~\eqref{sch-eq} in 2D once we set the $z$ component of all the relevant vectorial quantities to zero and neglect them in our calculations. {In particular, for $\bu:=(\ru_x,\ru_y)\in\R^2$,  we have $\vec\ru:=\ru_y$ and $\bu=(u_x,\vec u)$.}

Because $v$ is a short-range potential, bounded solutions $\psi(x,\vec\rr)$ of (\ref{sch-eq}) tend to superpositions of plane waves as $x\to\pm\infty$. This means that they admit asymptotic expressions of the form
    \be
    \frac{1}{(2\pi)^{\frac{3d+1}{2}}}
	\int_{\sD_k} \frac{d{\vec p}}{\varpi({\vec p})}\, e^{i{\vec p}\cdot{\vec r}}	\left[A_\pm({\vec p})\,e^{i\varpi({\vec p})x}+B_\pm({\vec p})\, e^{-i\varpi({\vec p})x}\right]~~\for~~x\to\pm\infty,
    \label{asym}
	\ee
where 
	\begin{align}
	&{\vec\rr:=\left\{\:\begin{array}{cc}
	y &\mbox{in 2D},\\
	(y,z) &\mbox{in 3D},
	\end{array}\right.}
	&&
	{\vec p:=\left\{\:\begin{array}{cc}
	p_y &\mbox{in 2D},\\
	(p_y,p_z) &\mbox{in 3D},
	\end{array}\right.}\\[3pt]	
	&\sD_k:=\left\{\:{\vec p}\in\R^2\,|\,{\vec p}^{\,2}< k^2\,\right\},
	&&\varpi({\vec p}):=\left\{\begin{array}{ccc}\sqrt{k^2-{\vec p}^{\,2}}&\for&{\vec p}\in\sD_k,\\ i\sqrt{{\vec p}^{\,2}-k^2}&\for&{\vec p}\notin\sD_k,\end{array}\right.
    \label{varpi=}
    \end{align}
and $A_\pm$ and $B_\pm$ are coefficient functions such that $A_\pm(\vec p)=B_\pm(\vec p)=0$ for $\vec p\notin\sD_k$.\footnote{Throughput this article we use the term function also for a tempered distribution.} We can state this condition as $A_\pm,B_\pm\in\sF_k^1$, where for all $m\in\Z^+$,
	\[\sF_k^m:=\left\{\,\bF\in\sF^m|~\bF({\vec p})=0~\for~{\vec p}\notin\sD_k\,\right\},\]
$\sF^m$ is the set of $m$-component functions $\bF:\R^d\to\C^{m\times 1}$, and $\C^{m\times n}$ denotes the vector space of $m\times n$ complex matrices. 
%In particular, $\sF^m=\sF^1\otimes\C^{m\times 1}$ and $\sF_k^m=\sF_k^1\otimes\C^{m\times 1}$. 
%We also introduce the projection operator $\widehat\Pi_k:\sF^m\to\sF^m$ that maps $\sF^m$ onto $\sF^m_k$. This is given by
%	\be
%	(\widehat\Pi_k \bF)({\vec p}):=\left\{\begin{array}{ccc}
%	\bF({\vec p}) & \for & {\vec p}\in\sD_k, \\
%	0 &  \for & {\vec p}\notin\sD_k.
%	\end{array}\right.
%	\label{project}
%	\ee

The fundamental transfer matrix is the linear operator $\widehat\bM:\sF_k^2\to\sF_k^2$ that satisfies 
	\be 
	\left[\begin{array}{c}
	A_+ \\ B_+ \end{array}\right]=\widehat{\bM}\left[\begin{array}{c}
	A_- \\ B_- \end{array}\right].
	\label{M-def}
	\ee
This is a direct generalization of the defining relation for the transfer matrix $\bM$ in one dimension, namely (\ref{M-1d}). Notice, however, that the entries $\widehat M_{ij}$ of $\widehat\bM$ are linear operators acting in $\sF_k^1$ which is an infinite-dimensional function space. As we show below, they are integral operators determined by the potential $v$.

Because the coefficient functions $A_\pm$ and $B_\pm$ determine the asymptotic behavior of the solutions of the Schr\"odinger equation (\ref{sch-eq}) and $\widehat\bM$ describes how they relate, it is not surprising to find out that $\widehat\bM$ determines the scattering amplitude of the potential. To see this, first we use the superscripts ``${\:l\:}$'' and ``${\:r\:}$'' to label the amplitude $N$ and the coefficient functions $A_\pm$ and $B_\pm$ for left- and right-incident waves, respectively.

For scattering solutions of (\ref{sch-eq}) corresponding to a left-incident wave, $\rn_{0x}>0$ and the term proportional to $A_-$ in \eqref{asym} represents the incident plane wave while the one proportional to $B_+$ must be absent.\footnote{This is because there is no source at $x=+\infty$ that could emit a wave traveling towards $x=0$.} More specifically, we have
	\begin{align}
	&A^l_-(\vec p)=N^l\check\delta_{\,\vec\rk_0}(\vec p), &&B^l_+(\vec p)=0,
	\label{scat-L}
	\end{align}
where for all ${\vec p\:}' \in\sD_k$,
	\be
	\check{\delta}_{{\vec p\:}'}(\vec p):=(2\pi)^d\varpi({\vec p\:}')\,\delta^d(\vec p-{\vec p\:}'),
	\label{chech-delta}
	\ee
$\vec \rk_0$ is the projection of the incident wave vector $\bk_0$ onto the $y$ axis in 2D and the $y$-$z$ plane in 3D, and $\delta^d(\cdot)$ stands for the Dirac delta function in $d$ dimensions. Employing Dirac's bra-ket notation, we can express \eqref{chech-delta} as
	\be
	|\check\delta_{{\vec p\:}'}\kt:=(2\pi)^d\varpi({\vec p\:}')\,|{\vec p\:}'\kt.
	\label{chech-delta-Dirac}
	\ee

The coefficient functions $B_-$ and $A_+$ appearing in \eqref{asym} respectively correspond to the waves reaching the detectors positioned at $x=-\infty$ and $x=+\infty$. For a left-incident wave, these are respectively the waves that are reflected back towards the source and the superposition of the incident wave and the scattered wave transmitted through the interaction region. Substituting \eqref{scat-L} in \eqref{asym} and comparing the result with \eqref{asymptot}, we find \cite{pra-2021}:
	\begin{align}
	&A^l_+(\vec\rk)=N^l\left[\check\delta_{\,\vec\rk_0}(\vec\rk)+c_d\,\ff(\bn_0,\bn)\right]&&\hspace{-3cm} \for~~~\rn_{0x}>0~{\rm and}~\rn_{x}>0, 
	\label{ALp=}\\
	&B^l_-(\vec\rk)=N^l c_d\,\ff(\bn_0,\bn)&&
	\hspace{-3cm} \for~~~\rn_{0x}>0~{\rm and}~\rn_x<0,
	\label{BLm=}
	\end{align}
where
	\be
	c_d:=(2\pi i)^{\frac{d}{2}}k^{1-\frac{d}{2}}.
	\label{cbk}
	\ee
Similarly, for a right-incident plane wave, where $\rn_{0x}<0$, we have
	\begin{align}
	&A^r_-(\vec p)=0,\quad\quad\quad\quad\quad\quad\quad 
	B^r_-(\vec p)=N^r\check\delta_{\,\vec\rk_0}(\vec p),
	\label{scat-R}\\
	&A^r_+(\vec\rk)=N^r c_d\,\ff(\bn_0,\bn)
	\quad \for~~~\rn_{0x}<0~{\rm and}~\rn_x>0,
	\label{ARp=}\\
	&B^r_-(\vec\rk)=N^r\left[\check\delta_{\,\vec\rk_0}(\vec\rk)+c_d\,\ff(\bn_0,\bn)\right]\quad \for~~~\rn_{0x}<0~{\rm and}~\rn_x<0.
	\label{BRm=}
	\end{align}
	
Next, we apply \eqref{M-def} to relate $A^{l/r}_\pm$ and $B^{l/t}_\pm$. In light of (\ref{scat-L}) and \eqref{scat-R}, this gives
	\begin{align}
	&\widehat M_{22}B_-^l=-N^l\widehat M_{21}\check\delta_{{\vec\rk}_0},
	&&\widehat M_{22}B_-^r=N^r \check\delta_{{\vec\rk}_0},
	\label{BLRm=}\\
	&A^{l}_+=N^l\widehat M_{11}\check\delta_{{\vec\rk}_0}+
	\widehat M_{12}B_-^{l},
	&&A^{r}_+=N^r\widehat M_{12}B_-^{r}.
	\label{ALRp=}
	\end{align}
We can formally express the solutions of (\ref{BLRm=}) as
	\begin{align}
	&B_-^l=-N^l\widehat M_{22}^{-1}\widehat M_{21}\check\delta_{{\vec\rk}_0},
	&&B_-^r=N^r\widehat M_{22}^{-1}\check\delta_{{\vec\rk}_0}.
	\label{Bs=}
	\end{align}
Substituting these in \eqref{ALRp=}, we obtain
	\begin{align}
	&A_+^l=N^l(\widehat M_{11}-\widehat M_{12}\widehat M_{22}^{-1}\widehat M_{21})\,\check\delta_{{\vec\rk}_0},
	&&A_+^r=N^r\widehat M_{12}\widehat M_{22}^{-1}\check\delta_{{\vec\rk}_0}.
	\label{As=}
	\end{align}
We can use (\ref{ALp=}), (\ref{BLm=}), (\ref{ARp=}), (\ref{BRm=}), \eqref{Bs=} and \eqref{As=} to express the scattering amplitude in terms of the entries of $\widehat\bM$. This gives
%	\be
%	\ff(\bn_0,\bn)=\frac{1}{c_d}\times\left\{\begin{array}{ccc}
%	\big((\widehat M_{11}-\widehat M_{12}\widehat M_{22}^{-1}\widehat M_{21}-\widehat I)\check\delta_{{\vec\rk}_0}\big)({\vec\rk}) &\for&\rn_{0x}>0~{\rm and}~\rn_x>0,\\[6pt]
%	-(\widehat M_{22}^{-1}\widehat M_{21}\check\delta_{{\vec\rk}_0})({\vec\rk}) 
%	&\for&\rn_{0x}>0~{\rm and}~\rn_x<0,\\[6pt]
%	(\widehat M_{12}\widehat M_{22}^{-1}\check\delta_{{\vec\rk}_0})({\vec\rk}) 
%	&\for&\rn_{0x}<0~{\rm and}~\rn_x>0,\\[6pt]
%	[(\widehat M_{22}^{-1}-\widehat I)\check\delta_{{\vec\rk}_0}]({\vec\rk}) 
%	&\for&\rn_{0x}<0~{\rm and}~\rn_x<0,
%	\end{array}\right.
%	\label{ff=}
%	\ee
	\be
	\ff(\bn_0,\bn)=\frac{(2\pi)^d\varpi(\vec \rk_0)}{c_d}\times\left\{\begin{array}{ccc}
	\br\vec\rk|(\widehat M_{11}-\widehat M_{12}\widehat M_{22}^{-1}\widehat M_{21}-\widehat I)|\vec\rk_0\kt
	&\for&\rn_{0x}>0~{\rm and}~\rn_x>0,\\[6pt]
	-\br\vec\rk|\,\widehat M_{22}^{-1}\widehat M_{21}|\vec\rk_0\kt
	&\for&\rn_{0x}>0~{\rm and}~\rn_x<0,\\[6pt]
	\br\vec\rk|\,\widehat M_{12}\widehat M_{22}^{-1}|\vec\rk_0\kt
	&\for&\rn_{0x}<0~{\rm and}~\rn_x>0,\\[6pt]
	\br\vec\rk|\,(\widehat M_{22}^{-1}-\widehat I)|\vec\rk_0\kt
	&\for&\rn_{0x}<0~{\rm and}~\rn_x<0,
	\end{array}\right.
	\label{ff=}
	\ee
where $\widehat I$ is the identity operator, and we have employed \eqref{chech-delta-Dirac} to express the final result in Dirac's bra-ket notation.\footnote{Note that if $\widehat L$ is a linear operator acting in $\sF_k^1$ and $\vec p\in\sD_k$, $\br \vec p|\widehat L f\kt:=(\widehat L f)(\vec p)$ and $\br\vec p|\widehat L|\vec p^{\,'}\kt$ is the integral kernel of $\widehat L$, which satisfies $(\widehat L f)(\vec p)=\int_{\sD_k} d^dp'\:\br\vec p|\widehat L|\vec p{\,'}\kt f(\vec p\,')$.} 

If zero belongs to the spectrum of $\widehat M_{22}$ for some $k\in\R^+$, $\ff(\bn_0,\bn)$ develops a singularity. This marks a spectral singularity \cite{prl-2009}. Because the intensity of the scattered wave is proportional to $|N^{l/r}\ff(\bn_0,\bn)|^2$, at the vicinity of a spectral singularity, the scattered wave attains a sizable intensity even for incident waves of arbitrarily small amplitude $N^{l/r}$.  This corresponds to a situation in which the system begins amplifying background noise and emitting coherent radiation, a phenomenon that is realized in every laser \cite{pra-2011,longhi-2010,ramezani-2014,SS-review}. 

Equation~\eqref{ff=} reduces the solution of the scattering problem for the potential $v$ to the determination of its fundamental transfer matrix. A highly nontrivial property of the latter is that, similarly to its one-dimensional analog, it can be expressed in terms of the evolution operator for a non-unitary effective quantum system. More specifically, it admits a Dyson series expansion of the form \cite{pra-2021}:
	\be
	\widehat\bM=\sT\exp\left[-i\int_{-\infty}^\infty dt~
	\widehat\bH(t)\right],
	\label{M=}
	\ee
where $\widehat{\bH}(x):\sF^2\to\sF^2$ is the effective Hamiltonian operator given by
	\begin{align}
	&\widehat\bH(x):=\frac{1}{2} e^{-ix\widehat{\varpi}_r
	\widehat{\bsigma}_3}	 \widehat{\sV}(x)\,\widehat{\varpi}^{-1} \widehat{\boldsymbol{\cK}}\,
	e^{ix\widehat{\varpi}_r\widehat{\bsigma}_3}
	-i\widehat{\varpi}_i\widehat{\bsigma}_3,
	\label{H=}
	\end{align}
$\widehat\varpi_r,\widehat\varpi,\widehat\varpi_i,\widehat{\bsigma}_3,\widehat\bcK,\widehat\sV(x):\sF^2\to\sF^2$ are linear operators defined by
    	\begin{align}
    	&(\widehat{\varpi}_r\bF)({\vec p}):=\RE[\varpi({\vec p})]\,\bF({\vec p})=
	\left\{\begin{array}{cc}
	\sqrt{k^2-\vec p^{\,2}}\:\bF(\vec p)&\for~|\vec p|<k,\\
	0&\for~|\vec p|\geq k,\end{array}\right.
	\label{pir}\\
	&(\widehat{\varpi}\bF)({\vec p}):= \varpi({\vec p})\bF({\vec p}),\quad\quad\quad 
	\widehat{\varpi}_i:=i(\widehat{\varpi}_r-\widehat{\varpi}),
	\quad\quad\quad 
	\big(\widehat{\bsigma}_3\bF\big)(\vec p):=\bsigma_3\bF(\vec p),
	\label{pi-pii}\\[6pt]
    	&\big(\widehat\bcK\,\bF\big)(\vec p):=\bcK\,\bF(\vec p), 
    	\quad\quad\quad
	\boldsymbol{\cK}:=\boldsymbol{\sigma}_3+i\boldsymbol{\sigma}_2=
	\left[\begin{array}{cc} 
	1 & 1\\
	-1 & -1\end{array}\right],
	\label{K-def}\\[6pt]
	&\big(\widehat\sV(x)\bF\big)({\vec p}):=\frac{1}{(2\pi)^d}\int_{\R^d} d^d{q}
    ~\tilde v(x,{\vec p}-{\vec q})\bF({\vec q}),
    \label{V=}
   	\end{align}
$\tilde v(x,{\vec p})$ stands for the (partial) Fourier transform of $v(x,{\vec r})$ with respect to ${\vec r}$, i.e., $\tilde v(x,{\vec p}):=$\linebreak $\int_{\R^d} d^dr\:e^{-i{\vec p}\cdot{\vec r}}v(x,{\vec r})$, and $\bsigma_j$ with $j\in\{1,2,3\}$ are the Pauli matrices;
	\begin{align}
	&\bsigma_1:=\left[\begin{array}{cc}
	0 & 1\\
	1 & 0\end{array}\right],
	&&\bsigma_2:=\left[\begin{array}{cc}
	0 & -i\\
	i & 0\end{array}\right],
	&&\bsigma_3:=\left[\begin{array}{cc}
	1 & 0\\
	0 & -1\end{array}\right].\nn
	\end{align}
Although $\widehat\bH(x)$ and consequently the right-hand side of \eqref{M=} are operators acting in $\sF^2$, one can show that the latter maps $\sF^2_k$ to $\sF^2_k$, \cite{pra-2021}. This allows us to view $\widehat\bM$ as an operator acting in $\sF^2_k$. Therefore, \eqref{M=} is consistent with the defining relation of $\widehat\bM$, namely \eqref{M-def}.

According to \eqref{M=} -- \eqref{V=}, $\widehat \sV(x)$, $\widehat\bH(x)$, and consequently $\widehat\bM$ and its entries $\widehat M_{ij}$ are linear integral operators. Reference~\cite{pra-2021} offers various examples where the latter can be computed analytically and used to obtain the exact solution of the corresponding scattering problems. {Among notable applications of the dynamical formulation of stationary scattering, which makes use of the concept of fundamental transfer matrix, are a singularity-free treatment of the delta-function point interactions in 2D and 3D  \cite{ap-2022} and the discovery of short-range potentials for which the Born approximation is exact. These results extend to electromagnetic scattering \cite{pra-2023,ptep-2024a}. Ref.~\cite{jpa-2024} uses the standard approach to stationary scattering based on the Lippmann-Schwinger equation to offer an independent confirmation of the results on the exactness of the $N$-th order Born approximation that were originally obtained using the fundamental transfer matrix.}

{We can view the effective Hamiltonian $\widehat\bH(x)$ as a linear operator acting in the space of square-integrable two-component wave functions $L^2(\R^{d})\otimes\C^{2\times 1}$. It is easy to see that in general this operator fails to be Hermitian. This is also true when $v$ is a real potential. In this case the non-Hermiticity of $\widehat\bH(x)$ does not conflict with the unitarity of the time-evolution of the original system, because the latter is determined by the Schr\"odinger operator $-\nabla^2+v(\bfr)$ which is a Hermitian operator acting in the Hilbert space $L^2(\R^{d+1})$ of the original system.}

 \section{Statement of the reciprocity relation in terms of $\widehat\bM$} 
	
The physical meaning of the coefficients functions $A^{l/r}_+$ and $B^{l/r}_-$ suggests defining the left and right reflection and transmission amplitudes according to
	\begin{align}
	&R^l(\bn_0,\bn):=\frac{k^{d-1}B^l_-({\vec\rk})}{N^l}=c_d\,k^{d-1}\ff(\bn_0,\bn) 
	&&\hspace{-1.5cm} \for~~~\rn_{0x}>0~{\rm and}~\rn_x<0,
	\label{RL-def}\\
	&T^l(\bn_0,\bn):=\frac{k^{d-1}A^l_+({\vec\rk})}{N^l}=k^{d-1}[\check\delta_{\,\vec\rk_0}({\vec\rk})+c_d\,\ff(\bn_0,\bn)]&&\hspace{-1.5cm} \for~~~\rn_{0x}>0~{\rm and}~\rn_x>0,
	\label{TL-def}\\
	&R^r(\bn_0,\bn):=\frac{k^{d-1}A^r_+({\vec\rk})}{N^r}=c_d\,k^{d-1}\ff(\bn_0,\bn)
	&&\hspace{-1.5cm}\for~~~\rn_{0x}<0~{\rm and}~\rn_x>0,
	\label{RR-def}\\
	&T^r(\bn_0,\bn):=\frac{k^{d-1}B^r_-({\vec\rk})}{N^r}=
	k^{d-1}[\check\delta_{\,\vec\rk_0}({\vec\rk})+c_d\,\ff(\bn_0,\bn)]&&\hspace{-1.5cm}\for~~~\rn_{0x}<0~{\rm and}~\rn_x<0,
	\label{TR-def}
	\end{align}
 where we have made use of \eqref{ALp=}, \eqref{BLm=}, \eqref{ARp=}, and \eqref{BRm=} and inserted the factor $k^{d-1}$ to ensure that $R^{l/r}$ and $T^{l/r}$ are dimensionless. See also \cite{nieto-vesperinas,scipost-2022}.
 %\footnote{$R^{l/r}$ and $T^{l/r}$ as defined by \eqref{RL-def} -- \eqref{TR-def} has the dimension of $k^{1-d}$. We could rescale their definition by a constant multiple of $k^{1-d}$ to render them dimensionless. We refrain from doing so, as it makes some of the formulas slightly more complicated.}  
 Substituting \eqref{ff=} in \eqref{RL-def} -- \eqref{TR-def}, we obtain 
 	\begin{align}
	&R^l(\bn_0,\bn)=-(2\pi)^d k^{d-1}\varpi(\vec\rk_0)\,\br\vec\rk|\,\widehat M_{22}^{-1}\widehat M_{21}|\vec\rk_0\kt 
	%&&\hspace{-1cm} \for~~~\rn_{0x}>0~{\rm and}~\rn_x<0
	,
	\label{RL=d}\\
	&T^l(\bn_0,\bn):=(2\pi)^d k^{d-1}\varpi(\vec\rk_0)\,\br\vec\rk|(\widehat M_{11}-\widehat M_{12}\widehat M_{22}^{-1}\widehat M_{21})|\vec\rk_0\kt 
	%&&\hspace{-1cm} \for~~~\rn_{0x}>0~{\rm and}~\rn_x>0
	,
	\label{TL=d}\\
	&R^r(\bn_0,\bn):=(2\pi)^d k^{d-1}\varpi(\vec\rk_0)\,\br\vec\rk|\,\widehat M_{12}\widehat M_{22}^{-1}|\vec\rk_0\kt 
	%&&\hspace{-1cm}\for~~~\rn_{0x}<0~{\rm and}~\rn_x>0
	,
	\label{RR=d}\\
	&T^r(\bn_0,\bn):=(2\pi)^d k^{d-1}\varpi(\vec\rk_0)\,\br\vec\rk|\,\widehat M_{22}^{-1}|\vec\rk_0\kt 
	%&&\hspace{-1cm}\for~~~\rn_{0x}<0~{\rm and}~\rn_x<0
	.
	\label{TR=d}
	\end{align}
These are $(d+1)$-dimensional analogs of (\ref{RT=}). We can also use (\ref{RL-def}) -- (\ref{TR-def}) to express the reciprocity condition (\ref{reciprocity-D23}) as the following constraints on the reflection and transmission amplitudes \cite{nieto-vesperinas}.
 	\begin{align}
	&R^{l}(\bn_0,\bn)=R^{l}(-\bn,-\bn_0)
	%~~~\for~~~\rn_{0x}>0~{\rm and}~\rn_x<0,
	,
	\label{RL-reciprocity}\\
	&R^{r}(\bn_0,\bn)=R^{r}(-\bn,-\bn_0)
	%~~~\for~~~	\rn_{0x}<0~{\rm and}~\rn_x>0,
	,
	\label{RR-reciprocity}\\
	&T^l(\bn_0,\bn)=T^r(-\bn,-\bn_0)
	%~~~\for~~~\rn_{0x}>0~{\rm and}~\rn_x>0
	,
	\label{TL-reciprocity}\\
	&T^r(\bn_0,\bn)=T^l(-\bn,-\bn_0)
	%~~~\for~~~\rn_{0x}<0~{\rm and}~\rn_x<0.
	.
	\label{TR-reciprocity}
	\end{align}
	
Let $\hat\bx$ be the unit vector pointing along the positive $x$ axis, and consider the special cases where $\bn_0,\bn\in\{-\hat\bx,\hat\bx\}$. Then, $\rn_{0x},\rn_{x}\in\{-1,1\}$, \eqref{RL-reciprocity} and \eqref{RR-reciprocity}  are trivially satisfied, and \eqref{TL-reciprocity} and \eqref{TR-reciprocity} coincide and give:
	\be
	T^l(\hat\bx,\hat\bx)=T^r(-\hat\bx,-\hat\bx).
	\label{T=T}
	\ee 
Recalling that the problem of finding the forward and backward scattering amplitudes for a potential that is a function of $x$ is equivalent to solving the scattering problem defined by the same potential in 1D, we can identify (\ref{T=T}) with (\ref{T=T1}). This shows that Theorem~1 follows as a corollary of Theorem~2.

Next, we address the question of whether the reciprocity relations \eqref{RL-reciprocity} -- \eqref{TR-reciprocity} can simplify the expression \eqref{TL=d} for 
$T^l$, as is the case in 1D. To see this, we introduce the reflection (parity) operator $\widehat{\cP}:\sF^m\to\sF^m$ defined by 
	\be
	(\widehat{\cP} \bF)(\vec p):=\bF(-\vec p).
	\label{cP-def}
	\ee
It is easy to see that it fulfills
	\begin{align}
	&\br\vec p|\widehat\cP=\br-\vec p|,
	&&\widehat\cP|\vec p\kt=|\!-\!\vec p\kt,
	&&\widehat{\cP}^2=\widehat I,
	&&[\widehat{\cP},\widehat\varpi]=\widehat 0,
	\label{id11}
	\end{align}
where $\hat I$ and $\hat 0$ are the identity and zero operators acting in $\sF^m$ (and $\sF^m_k$), and we have made use of \eqref{V=} and the identity: $\br\vec p|\vec p{\,'}\kt=\delta^d(\vec p-\vec p{\,'})=\br\vec p{\,'}|\vec p\kt$. Furthermore, in view of \eqref{varpi=}, \eqref{pi-pii}, and \eqref{id11}, we have 
	\be
	\varpi(-\vec p)\:|\!-\!\vec p\kt=\varpi(\vec p)\widehat\cP|\vec p\kt=
	\widehat\cP\varpi(\vec p)|\vec p\kt=
	\widehat\cP\widehat\varpi|\vec p\kt=
	\widehat\varpi\widehat\cP|\vec p\kt.
	\label{id12}
	\ee
{As we show in the appendix,} Eqs.~\eqref{RL=d} --  \eqref{TR=d}, \eqref{id11}, and \eqref{id12} allow us to identify the reciprocity relations \eqref{RL-reciprocity} -- \eqref{TR-reciprocity} with
	\begin{align}
	&\br\vec\rk|\,\widehat M_{22}^{-1}\widehat M_{21}\widehat\varpi\,|\vec\rk_0\kt=
	\br\vec\rk_0|\,\widehat{\cP}\widehat M_{22}^{-1}\widehat M_{21}\widehat\varpi\widehat{\cP}\,|\vec\rk\kt,
	\label{id-2}\\[3pt]
	&\br\vec\rk|\,\widehat M_{12}\widehat M_{22}^{-1}\widehat\varpi\,|\vec\rk_0\kt=
	\br\vec\rk_0|\,\widehat{\cP}\widehat M_{12}\widehat M_{22}^{-1}\widehat\varpi\widehat{\cP}\,
	|\vec\rk\kt,
	\label{id-3}\\[3pt]
	&\br\vec\rk|\,(\widehat M_{11}-\widehat M_{12}\widehat M_{22}^{-1}\widehat M_{21})\widehat\varpi\,|\vec\rk_0\kt=
	\br\vec\rk_0|\,\widehat{\cP}\widehat M_{22}^{-1}\widehat\varpi\widehat{\cP}\,|\vec\rk\kt,
	\label{id-4}
	\end{align}
where $\vec\rk$ and $\vec\rk_0$ are arbitrary elements of $\sD_k$. Substituting (\ref{id-4}) in \eqref{TL=d} and noting that $\widehat\varpi|\vec \rk\kt=\varpi(\vec \rk)|\vec \rk\kt$, we arrive at the following $(d+1)$-dimensional generalization of \eqref{Reciprocity-1D}.
	\be
	T^l(\bn_0,\bn)=(2\pi)^d k^{d-1}\varpi(\vec \rk)\,\br\vec\rk_0|\:
	\widehat{\cP}\widehat M_{22}^{-1}\widehat{\cP}\:|\vec\rk\kt
	%~~~\for~~~\rn_{0x}>0~{\rm and}~\rn_x>0.
	.
	\label{TL=d2}
	\ee

According to \eqref{TR=d} and \eqref{TL=d2}, the transmission properties of the potential is governed by the operator $\widehat M_{22}$. In particular, the potential enjoys perfect omnidirectional transparency if and only if $k$ is such that $\widehat M_{22}=\widehat I$. It possesses directional transparency along $\bn_0$ provided that
	\be
	\begin{aligned}
	&\widehat M_{22}^\dagger|-\vec\rk_0\kt=|-\vec\rk_0\kt
	&&\for~~\rn_{0x}>0,\\
	&\widehat M_{22}|\vec\rk_0\kt=|\vec\rk_0\kt
	&&\for~~\rn_{0x}<0.
	\end{aligned}
	\label{directional-T}
	\ee
Similarly, we can use \eqref{RL=d}, \eqref{RR=d}, and \eqref{id-3} to infer that the potential is omnidirectionally reflectionless if $\widehat M_{12}=\widehat M_{21}=\widehat 0$, and reflectionless along $\bn_0$ if and only if
	\be
	\begin{aligned}
	&\widehat M_{21}|\vec\rk_0\kt=0
	&&\for~~\rn_{0x}>0,\\
	&\widehat M_{12}^\dagger|-\vec\rk_0\kt=0
	&&\for~~\rn_{0x}<0.
	\end{aligned}
	\label{directional-R}
	\ee
	
%An interesting phenomenon happens when for some wavenumber $k$, the operator $\widehat M_{22}$ acquires a zero mode, i.e., there is some $|B_\star\kt$ such that $\widehat M_{22}|B_\star\kt=0$. Then, in view of \eqref{TR=d}, and \eqref{TL=d2}, there are some $|\vec\rk_0\kt$ and $|\vec\rk\kt$, and consequently $|\bn_0\kt$ and $|\bn\kt$, such that $T^r(\bn_0,\bn)$ and $T^l(-\bn_0,-\bn)$ blow up. This corresponds to the emergence of a spectral singularity \cite{prl-2009} at vicinity of which the system begins amplifying background noise and emitting coherence radiation, the phenomenon which is a generic feature of lasers \cite{pra-2011,longhi-2010,ramezani-2014,SS-review,konotop-2017}. 
	
Next, consider the antilinear operator $\widehat \cT:\sF^m\to\sF^m$ given by 
	\be
	(\widehat \cT \bF)(\vec p):=\bF(\vec p)^*,
	\label{cT-def}
	\ee
which satisfies
	\begin{align}
	&\br\vec p\,| \widehat \cT|f\kt=\br \vec p\,|f\kt^*,
	&& \widehat \cT|\vec p\,\kt=|\vec p\,\kt,
	&&\widehat \cT^2=\widehat I,
	&&[\widehat \cT,\widehat\varpi]=\widehat 0.
	\label{T-ids}
	\end{align}
	
It is easy to see that $\sF^m_k$ is an invariant subspace of $\sF^m$ for both $\widehat\cP$ and $\widehat\cT$. In fact their restrictions to $\sF^m_k$ define linear operators mapping $\sF^m_k$ onto $\sF^m_k$. We will use the same symbol for the restrictions of  $\widehat\cP$ and $\widehat\cT$ to $\sF^m_k$. 

If we view $\widehat\cP$, $\widehat\cT$, and $\widehat M_{ij}$ as operators acting in the Hilbert space $L^2(\sD_k)$ of square-integrable functions $f:\sD_k\to\C$, then $\widehat\cP$ is a Hermitian and unitary linear operator while $\widehat\cT$ is a Hermitian and antiunitary antilinear operator. It is also easy to check that 	\begin{align}
	&[\widehat\cP,\widehat\cT]=\widehat 0,
	&&(\widehat\cP\widehat\cT)^2=\widehat I,
	&&\br\vec p\,|\,\widehat \cT\, \widehat O^\dagger \widehat \cT\,|{\vec p}{\,'}\kt=
	\br\vec p|\widehat O^\dagger |\vec p{\,'}\kt^*=\br\vec p{\,'}|\widehat O|\vec p\kt,
	\label{id-15}
	\end{align}
where $\hat O:L^2(\sD_k)\to L^2(\sD_k)$ is any densely-define linear operator, and $\hat O^\dagger$ denotes its adjoint. {In the appendix we use (\ref{id-15}) to express the reciprocity relations (\ref{id-2}) -- \eqref{id-4} as the following conditions on the entries of the fundamental transfer matrix.}
	\begin{align}
	&(\widehat M_{22}^{-1}\widehat M_{21}\widehat\varpi)^\dagger=
	\widehat\cP\widehat\cT\,(\widehat M_{22}^{-1}\widehat M_{21}\widehat\varpi)\,\widehat\cP\widehat\cT,
	\label{id-2n}\\[3pt]
	&(\widehat M_{12}\widehat M_{22}^{-1}\widehat\varpi)^\dagger=
	\widehat\cP\widehat\cT\,(\widehat M_{12}\widehat M_{22}^{-1}\widehat\varpi)\,\widehat\cP\widehat\cT,
	\label{id-3n}\\[3pt]
	&(\widehat M_{22}^{-1}\widehat\varpi)^\dagger=\widehat\cP\widehat\cT\,
	(\widehat M_{11}-\widehat M_{12}\widehat M_{22}^{-1}\widehat M_{21})\widehat\varpi\,
	\widehat\cP\widehat\cT.
	\label{id-4n}
	\end{align}
	
Because $(\widehat\cP\widehat\cT)^{-1}=\widehat\cP\widehat\cT$, (\ref{id-2n}) and \eqref{id-3n} state that $\widehat M_{22}^{-1}\widehat M_{21}\widehat\varpi$ and $\widehat M_{12}\widehat M_{22}^{-1}\widehat\varpi$ are $\widehat\cP\widehat\cT$-anti-pseudo-Hermitian \cite{p3,paper183}. 
%Because $\widehat\cP\widehat\cT$ is an antiunitary operator, these operators have the same type of anti-pseudo-Hermiticity that is linked with reciprocal scattering in Ref.~\cite{Deak-Fulop}. 
{In the appendix, we use the fact that $\widehat\varpi$ and $\widehat\varpi^{-1}$ act in $L^2(\sD_k)$ as linear Hermitian operators to} express (\ref{id-2n}) -- \eqref{id-4n} in the form
	\begin{align}
	&(\widehat M_{22}^{-1}\widehat M_{21})^\dagger=
	\widehat\fT\,(\widehat M_{22}^{-1}\widehat M_{21})\,\widehat\fT^{-1},
	\label{id-2nn}\\[3pt]
	&(\widehat M_{12}\widehat M_{22}^{-1})^\dagger=
	\widehat\fT\,(\widehat M_{12}\widehat M_{22}^{-1})\,\widehat\fT^{-1},
	\label{id-3nn}\\[3pt]
	&\widehat M_{22}^{-1\dagger}=\widehat\fT\,
	(\widehat M_{11}-\widehat M_{12}\widehat M_{22}^{-1}\widehat M_{21})\,\widehat\fT^{-1},
	\label{id-4nn}
	\end{align}
where $\widehat\fT:L^2(\sD_k)\to L^2(\sD_k)$ is the Hermitian antilinear operator given by
	\be
	\widehat\fT:=\widehat\varpi^{-1}\widehat\cP\widehat\cT
	=\widehat\cP\widehat\cT\widehat\varpi^{-1}.
	\label{fT-def=}
	\ee
Equations \eqref{id-2nn} and \eqref{id-3nn} show that $\widehat M_{12}\widehat M_{22}^{-1}$ and $\widehat M_{12}\widehat M_{22}^{-1}$ are $\widehat\fT$-anti-pseudo-Hermitian operators \cite{p3,paper183}. Equation~\eqref{id-4nn} is equivalent to the following generalization of \eqref{det-M}.
	\be
	\widehat M_{11}\widehat M_{22}-\widehat M_{12}\widehat M_{22}^{-1}\widehat M_{21}\widehat M_{22}=\widehat\fT^{-1} \widehat M_{22}^{-1\dagger}\,\widehat\fT\,\widehat M_{22}.
	\label{det-M-d23}
	\ee
	
The above analysis proves the following result which we will use in the next section.
	\begin{itemize}
	\item[]{\bf Theorem~3} {\em Let $d\in\{1,2\}$ and $v:\R^{d+1}\to\C$ be a short-range potential with fundamental transfer matrix $\widehat\bM$. Then the reciprocity relation~\eqref{reciprocity-D23} holds if and only if the entries of $\widehat\bM$ satisfy \eqref{id-2nn} -- \eqref{id-4nn}.}
	\end{itemize}

\section{The property of $\widehat\bM$ that is responsible for reciprocity}

Equations~\eqref{id-2nn} -- \eqref{id-4nn} and \eqref{det-M-d23} give certain properties of the entries of the fundamental transfer matrix that are equivalent to the reciprocity condition \eqref{reciprocity-D23}. In Ref.~\cite{prsa-2016} we argue that the latter condition imposes a particular algebraic constraint on the fundamental transfer matrix. In this section, we offer a precise statement of this condition as an operator identity, establish its general validity, and give a proof of Theorem~2.

First, we reconsider the equivalence of the reciprocity condition \eqref{T=T1} in 1D and the requirement that the transfer matrix $\bM$ has unit determinant. The latter means that $\bM$ belongs to the special linear group $SL(2,\C)$. Because $SL(2,\C)$ is equal to the symplectic group $Sp(2,\C)$, we can identify the reciprocity condition \eqref{T=T1} with the requirement that $\bM\in Sp(2,\C)$. This means that 
	\be
	\bM^T\bOmega\,\bM=\bOmega,
	\label{symp-1}
	\ee
where the superscript ``\,$T$\,'' stands for the transpose of a matrix, and $\bOmega$ is the standard $2\times 2$ symplectic matrix;
	\be
	\bOmega:=\left[\begin{array}{cc}
	0 & 1\\
	-1 & 0\end{array}\right].
	\ee
	
Let $\cT:\C^{m\times 1}\to\C^{m\times 1}$ be the antiunitary operator of the complex-conjugations of $m\times 1$ complex matrices, i.e., for all $\bu\in\C^{m\times 1}$, $\cT(\bu)=\bu^*$. If we identify $2\times 2$ matrices $\bL$ with the linear operators $\bL:\C^{2\times 1}\to\C^{2\times 1}$ given by $\bL(\bu):=\bL\bu$ for all $\bu\in\C^{2\times 1}$, we can easily check that $\cT\,\bL^T\,\cT=\bL^{T*}=\bL^\dagger$. This observation allows us to identify \eqref{symp-1} and consequently the reciprocity condition \eqref{T=T1} with 
	\be
	(\bOmega\cT)^{-1}\bM^\dagger(\bOmega\cT)\,\bM=\bI,
	\label{ps-unitary1a}
	\ee
or
	\be
	\bM^\dagger=\bOmega\cT\,\bM^{-1}(\bOmega\cT)^{-1},
	\label{ps-unitary1}
	\ee
where we have employed the identity: $\bOmega^*=\bOmega=-\bOmega^{-1}$. Generalizing the terminology of Ref.~\cite{jmp-2004} for antilinear operators, we refer to (\ref{ps-unitary1a}) and (\ref{ps-unitary1}) as the $\bOmega\cT$-anti-pseudo-unitarity of $\bM$.

The main advantage of \eqref{ps-unitary1a} over \eqref{symp-1} is that by treating the matrices appearing in \eqref{ps-unitary1a} as linear operators, we can view the former as an operator identity. This is desirable, because it provides an expression for the reciprocity relation in 1D  that is valid in every matrix representation of these operators.\footnote{Equation~\eqref{symp-1} gives the statement of the reciprocity condition \eqref{T=T1} in the matrix representation defined by the standard basis of $\C^{2\times 1}$.} 

In what follows we obtain a higher-dimensional generalization of \eqref{ps-unitary1a} that is equivalent to the reciprocity condition~\eqref{reciprocity-D23}.

First, we consider a quantum system with Hilbert space $\sH$ and a possibly time-dependent Hamiltonian $\widehat H(t)$. Let $\widehat U(t,t_0):\sH\to\sH$ be the corresponding evolution operator, i.e., the linear operator satisfying 
	\begin{align}
	&i\partial_t\widehat U(t,t_0)=\widehat H(t)\widehat U(t,t_0),
	&&\widehat U(t_0,t_0)=\widehat I,
	\label{U-def}
	\end{align}
where $t,t_0\in\R$ are arbitrary, and $\widehat I$ is the identity operator acting in $\sH$.
	\begin{itemize}
	\item[] {\bf Lemma 1:} Let $t,t_0\in\R$, and $\widehat\cX:\sH\to\sH$ be a time-independent invertible antilinear operator. If $\widehat H(t)$ is $\widehat\cX$-pseudo-anti-Hermitian, i.e.,
	\be
	\widehat H(t)^\dagger=-\widehat\cX\,\widehat H(t)\,\widehat\cX^{-1},
	\label{ph-H}
	\ee
then  $\widehat U(t,t_0)$ is $\cX$-anti-pseudo-unitary, i.e.,
%	\be
	$\widehat U(t,t_0)^\dagger=\widehat\cX\,\widehat U(t,t_0)^{-1} \widehat\cX^{-1}$.
%	\label{pu-U}
%	\ee
\item[] Proof: Let $\widehat W(t,t_0):=\widehat\cX^{-1}\widehat U(t,t_0)^\dagger\widehat\cX\,\widehat U(t,t_0)$. Then the $\widehat\cX$-anti-pseudo-unitarity of $\widehat U(t,t_0)$ is equivalent to $\widehat W(t,t_0)=\widehat I$. It is easy to see that
	\begin{align}
	\partial_t \widehat W(t,t_0)&=\widehat\cX^{-1}[\partial_t\widehat U(t,t_0)^\dagger]\widehat\cX\,\widehat U(t,t_0)+\widehat\cX^{-1}\widehat U(t,t_0)^\dagger\widehat\cX\,\partial_t\widehat U(t,t_0)\nn\\
	&=-i\widehat\cX^{-1}\widehat U(t,t_0)^\dagger\widehat H(t)^\dagger\widehat\cX\,\widehat U(t,t_0)-i\widehat\cX^{-1}\widehat U(t,t_0)^\dagger\widehat\cX\,\widehat H(t)\widehat U(t,t_0)\nn\\
	&=\widehat 0,
	\label{dW=0}
	\end{align}
where $\widehat 0$ is the zero operator acting in $\sH$, and we have used \eqref{U-def} and (\ref{ph-H}). Because $\widehat W(t_0,t_0)=\widehat I$, (\ref{dW=0}) implies $\widehat W(t,t_0)=\widehat I$.~~~$\square$
\end{itemize}

Next, we view $\widehat\varpi_r$, $\widehat\varpi$, $\widehat\varpi_i$, $\widehat\bsigma_3$, $\widehat\bcK$, $\widehat\sV(x)$, $\widehat\cP$, and $\widehat\cT$ defined by \eqref{pir} -- \eqref{V=}, \eqref{cP-def}, and \eqref{cT-def} as linear or antilinear operators acting in the Hilbert space $L^2(\R^d)\otimes\C^{2\times 1}$, and let $\widehat\bOmega:L^2(\R^d)\otimes\C^{2\times 1}\to L^2(\R^d)\otimes\C^{2\times 1}$ be the linear operator given by
	\begin{align}
	&\big(\widehat\bOmega\bF\big)(\vec p):=\bOmega\,\bF(\vec p).\nn
	\end{align}
Then we can identify the Hamiltonian operator $\widehat\bH(x)$ given by \eqref{H=} and the fundamental transfer matrix $\widehat\bM$ as linear operators acting in $L^2(\R^d)\otimes\C^{2\times 1}$ and $L^2(\sD_k)\otimes\C^{2\times 1}$, respectively.
	
It is not difficult to see that the Hilbert space $L^2(\sD_k)\otimes\C^{2\times 1}$ is an invariant subspace of  $L^2(\R^d)\otimes\C^{2\times 1}$ for $\widehat\varpi$, $\widehat\cP$, $\widehat\cT$, and $\widehat\bOmega$, i.e., these operators and their inverses map $L^2(\sD_k)\otimes\C^{2\times 1}$  to $L^2(\sD_k)\otimes\C^{2\times 1}$. We use this fact to introduce the extension of the antilinear operator $\widehat\fT$ to $L^2(\sD_k)\otimes\C^{2\times 1}$. This is the antilinear Hermitian operator $\widehat\fT:L^2(\sD_k)\otimes\C^2\to L^2(\sD_k)\otimes\C^2$ defined by \eqref{fT-def=}, where we use the same symbols for $\widehat\varpi^{-1}$, $\widehat\cP$, $\widehat\cT$, and their restrictions to $L^2(\sD_k)\otimes\C^2$.

\begin{itemize}
	\item[]{\bf Theorem~4} {\em Let $d\in\{1,2\}$ and $v:\R^{d+1}\to\C$ 
	 be a short-range potential. Then the fundamental transfer matrix $\widehat\bM$ of $v$ viewed as a linear operator acting in $L^2(\sD_k)\otimes\C^2$ is $\widehat\bOmega\widehat\fT$-anti-pseudo-unitary, i.e.,
	it satisfies the following generalization of (\ref{ps-unitary1a}).}
	 \be
(\widehat\bOmega\,\widehat\fT)^{-1}\widehat\bM^\dagger
(\widehat\bOmega\,\widehat\fT)\,\widehat\bM=\widehat I.
	\label{ps-unitary1a-D23}
	\ee 
	\item[]Proof: As operators acting in $L^2(\R^d)\otimes\C^{2\times 1}$, 
	$\widehat\varpi_r$, $\widehat\varpi$, $\widehat\varpi_i$, $\widehat\cP$, $\widehat\cT$, $\widehat\sV(x)$, $\widehat\bOmega$, $\widehat\bcK$, and $\widehat\bsigma_3$ fulfill the following identities. 
	\begin{align}
	&\widehat\varpi_r^\dagger=\widehat\varpi_r=
	%\frac{1}{2}(\widehat\varpi+\widehat\varpi^\dagger)=
	\widehat\cT\,\widehat\varpi_r\,\widehat\cT=
	\widehat\cP\,\widehat\varpi_r\,\widehat\cP=
	\widehat\cP\widehat\cT\,\widehat\varpi_r\,(\widehat\cP\widehat\cT)^{-1},
	\label{eq-zz1}\\
	&\widehat\varpi_i^\dagger=\widehat\varpi_i=
	%\frac{1}{2i}(\widehat\varpi-\widehat\varpi^\dagger)=
	\widehat\cT\,\widehat\varpi_i\,\widehat\cT=
	\widehat\cP\,\widehat\varpi_i\,\widehat\cP=
	\widehat\cP\widehat\cT\,\widehat\varpi_i\,(\widehat\cP\widehat\cT)^{-1},
	\label{eq-zz2}\\
	&\widehat\varpi^\dagger=\widehat\cT\,\widehat\varpi\,\widehat\cT=
	\widehat\cT\,\widehat\cP\,\widehat\varpi\,\widehat\cP\,\widehat\cT=
	\widehat\cP\widehat\cT\,\widehat\varpi\,(\widehat\cP\widehat\cT)^{-1},
	\label{eq-zz3}\\%[3pt]
	&\widehat{\sV}(x)^\dagger=\widehat\cP\widehat\cT\,\widehat{\sV}(x)\,\widehat\cP\widehat\cT=
	\widehat\cP\widehat\cT\,\widehat{\sV}(x)\,(\widehat\cP\widehat\cT)^{-1},
	\label{eq-zz4}\\%[3pt]
	&[\widehat\varpi_r,\widehat\bOmega]=[\widehat\varpi_i,\widehat\bOmega]=
	[\widehat\varpi,\widehat\bOmega]=[\widehat{\sV}(x),\widehat\bOmega]=\widehat 0,
	\label{eq-zz45}\\
	&\widehat\bOmega\widehat\cT\,\widehat\bcK\,(\widehat\bOmega\widehat\cT)^{-1}=-\widehat\bcK^\dagger,
	\label{eq-zz5}\\
	&\widehat\bOmega\widehat\cT\,\widehat\bsigma_3\,(\widehat\bOmega\widehat\cT)^{-1}=-\widehat\bsigma_3=-\widehat\bsigma_3^\dagger.
	\label{eq-zz6}
	\end{align}
These together with \eqref{H=} and \eqref{fT-def=} imply
	\be
	\widehat\bOmega\,\widehat\fT\,\widehat\bH(x)(\widehat\bOmega\,\widehat\fT)^{-1}=-\widehat\bH(x)^\dagger,
	\label{H-anti-ph}
	\ee
i.e., $\widehat\bH(x)$ is $\widehat\bOmega\,\widehat\fT$-pseudo-anti-Hermitian. In light of Lemma~1, this implies that the evolution operator $\widehat\bU(x,x_0)$ corresponding to $\widehat\bH(x)$ is an $\widehat\bOmega\,\widehat\fT$-anti-pseudo-unitary operator. That is
	\be
(\widehat\bOmega\,\widehat\fT)^{-1}\widehat\bU(x,x_0)^\dagger
(\widehat\bOmega\,\widehat\fT)\,\widehat\bU(x,x_0)=\widehat I.
	\label{ps-unitary1-bU}
	\ee
Applying both sides of this equation to the elements of $L^2(\sD_k)\otimes\C^{2\times 1}$ (that belong to the domain of $\widehat\bM$) and taking their limits as $x\to+\infty$ and $x_0\to-\infty$, we are led to \eqref{ps-unitary1a-D23}.~~~$\square$
	\end{itemize}	

	\begin{itemize}
	\item[]{\bf Theorem~5} {\em Let $d\in\{1,2\}$ and $v:\R^{d+1}\to\C$ be a short-range potential with  fundamental transfer matrix $\widehat\bM$. Then the operator identity \eqref{ps-unitary1a-D23} is equivalent to the requirement that the entries of $\widehat\bM$ satisfy \eqref{id-2nn} -- \eqref{id-4nn}.}
	\item[]Proof: Expressing \eqref{ps-unitary1a-D23} in terms of the entries of $\widehat\bM$ we obtain the following three independent relations.
	\begin{align}
	&\widehat M_{11}^\dagger\widehat\fT\,\widehat M_{21}=	
	\widehat M_{21}^\dagger \widehat\fT\,\widehat M_{11} ,
	\label{id-01}\\
	&\widehat M_{22}^\dagger \widehat\fT\,\widehat M_{12}=	
	\widehat M_{12}^\dagger \widehat\fT\,\widehat M_{22},
	\label{id-04}\\
	&\widehat M_{22}^{\,\dagger} \widehat\fT\,\widehat M_{11}=
	\widehat M_{12}^{\,\dagger} \widehat\fT\,\widehat M_{21}+\widehat\fT.
	\label{id-03}
	\end{align}
Therefore, to prove this theorem, it is sufficient to show the equivalence of these equations to \eqref{id-2nn} -- \eqref{id-4nn}. To do this, first we write \eqref{id-04} in the form:
	\begin{align}
	\widehat M_{12}\widehat M_{22}^{-1}
	&=\widehat\fT^{-1}\widehat M_{22}^{-1\dagger} \widehat M_{12}^{\,\dagger} 
	\widehat\fT\nn\\
	&=\widehat\fT^{-1}(\widehat M_{12}\widehat M_{22}^{-1})^\dagger\,
	\widehat\fT,
	\label{id-04new}
	\end{align}
which is equivalent to \eqref{id-3nn}. Next, we express \eqref{id-03} as
	\begin{align}
	\widehat\fT\,\widehat M_{11}
	&=\widehat M_{22}^{-1\dagger} \widehat \fT+
	\widehat M_{22}^{-1\dagger}\widehat M_{12}^{\,\dagger}\,
	\widehat\fT\,\widehat M_{21}\nn\\
	&=\widehat M_{22}^{-1\dagger} \widehat \fT+
	\widehat M_{22}^{-1\dagger}(\widehat M_{12}^{\,\dagger}\,
	\widehat\fT\,\widehat M_{22})\widehat M_{22}^{-1}\widehat M_{21}\nn\\
	&=\widehat M_{22}^{-1\dagger} \widehat \fT+
	\widehat \fT\,\widehat M_{12}\widehat M_{22}^{-1}\widehat M_{21},
	\label{id-03a}
	\end{align}
where we have made use of \eqref{id-04}. Equation~\eqref{id-03a} is equivalent to \eqref{id-4nn}. In view of the equivalence of \eqref{id-04} and \eqref{id-3nn}, this establishes the equivalence of \eqref{id-04} and \eqref{id-03} to \eqref{id-3nn} and \eqref{id-4nn}. Next, we write \eqref{id-01} in the form
	\be
	(\widehat\fT\widehat M_{11})^\dagger
	\widehat M_{21}=
	\widehat M_{21}^\dagger\widehat\fT\widehat M_{11},
	\label{id-01z} 
	\ee
and express \eqref{id-03} as
	\be
	\widehat\fT\widehat M_{11}=\widehat M_{22}^{-1\dagger}(\widehat M_{12}^\dagger\widehat\fT\,\widehat M_{21}+\widehat\fT).\nn
	\ee
Substituting this relation in (\ref{id-01z}), we have
	\be
	\widehat M_{21}^\dagger\widehat M_{22}^{-1\dagger}\widehat\fT
	=\widehat\fT\,\widehat M_{22}^{-1}\widehat M_{21}
	+\widehat M_{21}^\dagger(\widehat\fT\,\widehat M_{12}
	\widehat M_{22}^{-1}-\widehat M_{22}^{-1\dagger}
	\widehat M_{12}^\dagger\widehat\fT)\widehat M_{21}.\nn
	\ee
We can write this equation in the form
	\begin{align}
	(\widehat M_{22}^{-1}\widehat M_{21})^\dagger&=
	\widehat\fT\,\widehat M_{22}^{-1}\widehat M_{21}\,\widehat\fT^{-1}+
	\widehat M_{21}^\dagger[
	\widehat\fT\,\widehat M_{12}\widehat M_{22}^{-1}
	\,\widehat\fT^{-1}-(\widehat M_{12}\widehat M_{22}^{-1})^\dagger)]
	\widehat\fT\widehat M_{21}\,\widehat\fT^{-1}\nn\\
	&=
	\widehat\fT\,\widehat M_{22}^{-1}\widehat M_{21}\,\widehat\fT^{-1},
	\label{id-01-new}
	\end{align}
where the last equation follows from \eqref{id-04new} which is equivalent to  \eqref{id-04}. Equation \eqref{id-01-new} is identical to \eqref{id-2nn}. Because \eqref{id-04} and \eqref{id-03} are equivalent to \eqref{id-2nn} and \eqref{id-4nn}, this establishes the equivalence of \eqref{id-01} -- \eqref{id-03} and \eqref{id-2nn} -- \eqref{id-4nn}.~~~$\square$
\end{itemize}

Theorems 3 and 5 imply the equivalence of the reciprocity relation~\eqref{reciprocity-D23} and the operator identity \eqref{ps-unitary1a-D23}. Theorem 4 shows that the latter is always satisfied. This provides a proof of the reciprocity theorem in 2D and 3D (Theorem 2), which in contrast to the earlier proofs of this theorem, does not make use of the properties of the scattering operator (S matrix) \cite{Landau}, those of the resolvent operators for the corresponding Schr\"odinger operators (Green's functions) \cite{Twersky1954,Deak-Fulop,Bilhorn-1964}, or Green's identities \cite{Masoud-2019,Wapenaar,Carminati}. It only relies on a basic property of the fundamental transfer matrix, namely its $\widehat\bOmega\widehat\fT$-anti-pseudo-unitarity \eqref{ps-unitary1a-D23}.

\section{Anti-pseudo-Hermiticity of the scattering matrix}

In 1D the scattering operator can be conveniently represented by a $2\times 2$ matrix that maps the amplitudes $A_-$ and $B_+$ of the incoming waves to the amplitudes $A_+$ and $B_-$ of the outgoing waves in the asymptotic expression \eqref{pw} for the solutions of the Schr\"odinger equation \eqref{sch-eq-1d}. Depending on how we arrange the pairs $(A_-,B_+)$ and $(A_+,B_-)$ into a $2\times 1$ matrix, we have four different ways of defining the scattering matrix, \cite{bookchapter}. The most popular choices are given by \cite{muga-review,Ge-2012}:
	\begin{align}
	&\bS \left[\begin{array}{c}
	A_-\\B_+\end{array}\right]:=\left[\begin{array}{c}
	A_+\\B_-\end{array}\right],
	&& \bS' \left[\begin{array}{c}
	A_-\\B_+\end{array}\right]:=\left[\begin{array}{c}
	B_-\\A_+\end{array}\right].\nn
	\end{align}
 Applying these equations for the left- and right-incident waves, we find
 	\begin{align}
	&\bS=\left[\begin{array}{cc}
	T^l&R^l\\
	R^r & T^r\end{array}\right]=\frac{1}{M_{22}}
	\left[\begin{array}{cc}
	\det\bM &-M_{21}\\
	M_{12} & 1\end{array}\right],
	\label{S=}
	\\
	&\bS'=\left[\begin{array}{cc}
	R^l&T^r\\
	T^l & R^r\end{array}\right]=\bsigma_1\bS,
	\label{S2=}
	\end{align}
where we have made use of \eqref{RT=}.

The reciprocity relation \eqref{T=T1} is equivalent to the assertion that $\bS'$ is a symmetric matrix. In view of the last equation in (\ref{S2=}) and the identities 
	\begin{align}
	&\bS^{'\dagger}=\cT\,\bS^{'T}\cT, &&\bS^\dagger=\cT\bS^T\,\cT, && \bsigma_1^{-1}=\bsigma_1=\cT\,\bsigma_1\cT, &&\cT^{-1}=\cT,\nn
	\end{align}
we can express \eqref{T=T1} as
	\be
	\bS^{'\dagger}=\cT\,\bS'\cT^{-1},
	\label{recirocity-S-prime}
	\ee
or
	\be
	\bS^\dagger=(\bsigma_1\cT)\bS(\bsigma_1\cT)^{-1}.
	\label{recirocity-S}
	\ee
Identifying $\bS$, $\bsigma_1$, and $\bS'$ with linear operators acting in $\C^{2\times 1}$ and the complex conjugation $\cT$ as an antilinear operator acting in $\C^{2\times 1}$, we can view  \eqref{recirocity-S-prime} and \eqref{recirocity-S} as operator identities that are equivalent to the reciprocity relation \eqref{T=T1}. They mean that $\bS'$ and $\bS$ are respectively $\cT$-anti-pseudo-Hermitian and $\bsigma_1\cT$-anti-pseudo-Hermitian. We obtain 2D and 3D generalizations of these identities in the sequel. 

First, we define the scattering matrix in $d+1$ dimensions as the linear operator $\widehat\bS:\sF^2_k\to\sF^2_k$ that satisfies 
	\begin{align}
	&\widehat\bS \left[\begin{array}{c}
	A_-\\B_+\end{array}\right]:=\left[\begin{array}{c}
	A_+\\B_-\end{array}\right],
	\label{S-def}
	\end{align}
where $A_\pm$ and $B_\pm$ are the coefficient functions determining the asymptotic expression \eqref{asym} for the bounded solutions of the Schr\"odinger equation \eqref{sch-eq}, \cite{scipost-2022}. Substituting the coefficient functions $A_\pm^{l/r}$ and $B_\pm^{l/r}$ for the left-/right-incident waves in \eqref{S-def} and employing \eqref{scat-L} -- \eqref{chech-delta-Dirac}, \eqref{scat-R}, and \eqref{RL-def} -- \eqref{TR-def}, we then find
	\begin{align}
	\br\vec\rk|\,\widehat\bS\,|\vec\rk_0\kt=\frac{k^{1-d}}{(2\pi)^2\varpi(\vec\rk_0)}
	\left[\begin{array}{cc}
	T^l(\bn_0,\bn) &R^r(\bn_0,\bn)\\
	R^l(\bn_0,\bn) & T^r(\bn_0,\bn)\end{array}\right].\nn
	\end{align}
In view of \eqref{RL=d} --	\eqref{TR=d}, we can express this equation as the following generalization of the last equation in \eqref{S=}.
	\begin{align}
	\widehat\bS=\left[\begin{array}{cc}
	\widehat\bM_{11}-\widehat\bM_{12}\widehat\bM_{22}^{-1}\widehat\bM_{21}	
	&~~~\widehat\bM_{12}\widehat\bM_{22}^{-1}\\[6pt]
	-\widehat\bM_{22}^{-1}\widehat\bM_{21} 
	&~~~\widehat\bM_{22}^{-1}\end{array}\right].
	\label{S=M}
	\end{align}
	
Next, we view $\widehat\bS$ as a linear operator acting in the Hilbert space $L^2(\sD_d)\otimes\C^{2\times 1}$, and introduce the linear operator $\widehat\bsigma_1:L^2(\sD_d)\otimes\C^{2\times 1}\to L^2(\sD_d)\otimes\C^{2\times 1}$ that is given by
	\begin{align}
	&\big(\widehat\bsigma_1\bF\big)(\vec p):=\bsigma_1\bF(\vec p).
	\end{align}
\begin{itemize}
	\item[]{\bf Theorem~6} {\em Let $d\in\{1,2\}$ and $v:\R^{d+1}\to\C$ be a short-range potential. Then the scattering matrix $\widehat\bS$ of $v$ that is given by \eqref{S-def} is $\widehat\bsigma_1\widehat\fT$-anti-pseudo-Hermitian, i.e., it satisfies the following generalization of \eqref{recirocity-S}.}
	\be
	\widehat\bS^\dagger=(\widehat\bsigma_1\widehat\fT)\,\widehat\bS\,(\widehat\bsigma_1\widehat\fT)^{-1}.
	\label{S-reciprocity}
	\ee
	\item[]Proof: First, we note that according to Theorems 4 and 5 the entries of the fundamental transfer matrix satisfy \eqref{id-2nn} -- \eqref{id-4nn}. These together with \eqref{S=M} and the identities, 
	\begin{align}
	&\widehat\bsigma_1^\dagger=\widehat\bsigma_1
	=\widehat\bsigma_1^{-1},
	&&[\widehat\bsigma_1,\widehat\fT]
	=[\widehat\bsigma_1,\widehat\fT^{-1}]
	=\widehat 0,
	\label{id-245}
	\end{align}	
imply
		\begin{align}
	\widehat\bsigma_1^{-1}\widehat\bS^\dagger\widehat\bsigma_1
	&=\widehat\bsigma_1\left[\begin{array}{cc}
	[\widehat\bM_{11}-\widehat\bM_{12}\widehat\bM_{22}^{-1}\widehat\bM_{21}]^\dagger
	&~~~-(\widehat\bM_{22}^{-1}\widehat\bM_{21})^\dagger\\[6pt]
	(\widehat\bM_{12}\widehat\bM_{22}^{-1})^\dagger	
	&~~~\widehat\bM_{22}^{-1\dagger}\end{array}\right]\widehat\bsigma_1\nn\\
	&=\widehat\bsigma_1\widehat\fT
	\left[\begin{array}{cc}
	\widehat\bM_{22}^{-1}~~~& -\widehat\bM_{22}^{-1}\widehat\bM_{21}\\[6pt]
	\widehat\bM_{12}\widehat\bM_{22}^{-1}~~~&
	\widehat\bM_{11}-\widehat\bM_{12}\widehat\bM_{22}^{-1}\widehat\bM_{21}
	\end{array}\right]\widehat\fT^{-1}\widehat\bsigma_1\nn\\[6pt]
	&=\widehat\fT\,\widehat\bS\,\widehat\fT^{-1}.\nn
	\end{align}
By virtue of \eqref{id-245}, the last equation is equivalent to \eqref{S-reciprocity}.~~~$\square$
\end{itemize}
	
Because \eqref{id-2nn} -- \eqref{id-4nn} are equivalent to the reciprocity condition~\eqref{reciprocity-D23}, the proof of Theorem~6 shows that the $\widehat\bsigma_1\widehat\fT$-anti-pseudo-Hermiticity of the scattering matrix $\widehat\bS$ is a consequence of the reciprocity or the $\widehat\bOmega\widehat\fT$-anti-pseudo-unitarity of the fundamental transfer matrix. We can also state a similar result for the scattering matrix,
	\be
	\widehat\bS':=\widehat\bsigma_1\widehat\bS=
	\left[\begin{array}{cc}
	-\widehat\bM_{22}^{-1}\widehat\bM_{21} 
	&~~~\widehat\bM_{22}^{-1}\\[6pt]
	\widehat\bM_{11}-\widehat\bM_{12}\widehat\bM_{22}^{-1}\widehat\bM_{21}	
	&~~~\widehat\bM_{12}\widehat\bM_{22}^{-1}
	\end{array}\right].
	\label{S-prime=M}
	\ee
It is easy to show that the $\widehat\bsigma_1\widehat\fT$-anti-pseudo-Hermiticity of $\widehat\bS$ is equivalent to the $\widehat\fT$-anti-Hermiticity of $\widehat\bS'$; \eqref{S-reciprocity}	holds if and only if
	\be
	\widehat\bS^{'\dagger}= \widehat\fT\,\widehat\bS'\,\widehat\fT^{-1}.
	\label{S-prime-reciprocity}
	\ee
This is the generalization of \eqref{recirocity-S-prime} to 2D and 3D.

\section{Concluding remarks}

Reciprocity is an important aspect of potential scattering. It was discovered in the nineteenth century, decades before the formulation of quantum scattering theory.  But its origins and practical implications have continued to attract the attention of physicists. The present article explores the theoretical roots of this phenomenon in the dynamical formulation of stationary scattering where the scattering of waves is related to the dynamics of an effective non-unitary quantum system. This is based on the notion of the fundamental transfer matrix which is a generalization to two and three dimensions of the well-known transfer matrix of potential scattering in one dimension.

The Hamiltonian operator of the effective quantum system that determined the fundamental transfer matrix is non-Hermitian. We have traced the origin of reciprocity in potential scattering to an operator identity satisfied by this Hamiltonian. This identity involves a particular potential-independent antilinear Hermitian operator that renders the fundamental transfer matrix anti-pseudo-unitary. We have shown that this feature of the fundamental transfer matrix is equivalent to the reciprocity relation, thus providing an alternative proof of the reciprocity theorem.

As by-products of our analysis we have uncovered 2D and 3D generalizations of a number of well-known formulas of potential scattering in one dimension and established a particular generic anti-pseudo-Hermiticity of the scattering operator that is linked with the anti-pseudo-unitarity of the fundamental transfer matrix and consequently the reciprocity relation. \vspace{12pt}

\noindent {\bf Data Availability:} No datasets were generated or analyzed during the current study.
\vspace{12pt}

\noindent {\bf Acknowledgements:}
This work has been supported by Turkish Academy of Sciences (T\"UBA).	

{\section*{Appendix: Derivation of Eqs.~\eqref{id-2} -- \eqref{id-4} and \eqref{id-2n} -- \eqref{id-4nn}}}

{First, we use \eqref{pi-pii}, \eqref{RL=d}, and \eqref{RL-reciprocity} to establish
    \be
    \label{Appendix-1}
    \br\vec\rk|\,\widehat M_{22}^{-1}\widehat M_{21}\widehat\varpi|\vec\rk_0\kt =\br-\,\vec\rk_0|\,\widehat M_{22}^{-1}\widehat M_{21}\widehat\varpi
    |\!-\vec\rk\kt.
    \ee
In view of \eqref{id11}, this equation implies \eqref{id-2}. Similarly, \eqref{RR=d} and \eqref{RR-reciprocity} imply that \eqref{Appendix-1} holds with $\widehat M_{22}^{-1}\widehat M_{21}$ changed to $\widehat M_{12}\widehat M_{22}^{-1}$. This proves \eqref{id-3}. Eq.~\eqref{id-4} follows from \eqref{TL=d}, \eqref{TR=d}, and \eqref{TL-reciprocity}, for we can use them to show that 
    \bea
    \br\vec\rk|(\widehat M_{11}-\widehat M_{12}\widehat M_{22}^{-1}\widehat M_{21})\widehat\varpi|\vec\rk_0\kt
    &=&\br-\vec\rk_0|\widehat M_{22}^{-1}\widehat\varpi|-\!\vec\rk\kt\nn\\ 
    &=&\br\vec\rk_0|\,\widehat{\cP} \widehat M_{22}^{-1}\widehat\varpi \widehat{\cP}\,|\vec\rk\kt.\nn
    \eea
This completes the derivation of \eqref{id-2} -- \eqref{id-4}.} 
   
{Next, we use \eqref{id-2} and \eqref{id-15} to show that 
	\begin{align}
	 \br\vec\rk_0|\,\widehat{\cT}\big(\widehat M_{22}^{-1}\widehat M_{21}\widehat\varpi\big)^\dagger\widehat{\cT}\,|\vec\rk\kt&=
	 \br\vec\rk_0|\,\widehat{\cP}\widehat M_{22}^{-1}\widehat M_{21}\widehat\varpi\widehat{\cP}\,|\vec\rk\kt.
	 \label{Appendix-2a}
	\end{align}
Because $\widehat{\cT}^2=\widehat{I}$ and $[\widehat{\cP},\widehat{\cT}]=\widehat 0$, \eqref{Appendix-2a} implies \eqref{id-2n}. Equation~\eqref{id-3n} follows similarly from \eqref{id-3} and \eqref{id-15}, as they imply \eqref{Appendix-2a} with $\widehat M_{22}^{-1}\widehat M_{21}$ changed to $\widehat M_{12}\widehat M_{22}^{-1}$.
To derive \eqref{id-4n}, we note that according to \eqref{id-4} and \eqref{id-15},
	\begin{align}
    	\br\vec\rk|(\widehat M_{11}-\widehat M_{12}\widehat M_{22}^{-1}\widehat M_{21})\,\widehat\varpi|\vec\rk_0\kt&=
	\br \vec\rk|\,\widehat\cT\big(\widehat\cP\widehat M_{22}^{-1}
	\widehat\varpi\widehat\cP\big)^\dag\widehat\cT\,|\vec\rk_0\kt.
	\label{Appendix-2b}
    \end{align}
Because $\widehat\cP^\dag=\widehat\cP$ and $(\widehat\cP\widehat\cT)^2=\widehat I$, \eqref{Appendix-2b} implies  \eqref{id-4n}. This completes the derivation of \eqref{id-2n} -- \eqref{id-4n}.}

{Equations~\eqref{id-2nn} -- \eqref{id-4nn} follow from \eqref{id-2n} -- \eqref{id-4n} if we multiply the left-hand sides of the latter equations by $\widehat\varpi^{-1}$ from the left and making use of \eqref{fT-def=} and the fact that $\widehat\fT^{-1}=\widehat\varpi\,\widehat\cP\widehat\cT$.}

{It is easy to see that the logical chain of arguments we use to derive \eqref{id-2nn} -- \eqref{id-4nn} from \eqref{RL-reciprocity} -- \eqref{TR-reciprocity} can be used in the reverse order to derive \eqref{RL-reciprocity} -- \eqref{TR-reciprocity} from \eqref{id-2nn} -- \eqref{id-4nn}. This proves the equivalence of the reciprocity relations \eqref{RL-reciprocity} -- \eqref{TR-reciprocity} and either of the triplets of equations \eqref{id-2} -- \eqref{id-4}, \eqref{id-2n} -- \eqref{id-4n}, and \eqref{id-2nn} -- \eqref{id-4nn}.}

\ed